\documentclass[11pt,a4paper]{article}
\pdfoutput=1
\usepackage{jheppub}
\usepackage{slashed}
\usepackage{amsmath}
\usepackage{amssymb}
\usepackage{epsfig}

\def\bea{\begin{eqnarray}}
\def\eea{\end{eqnarray}}
\def\nn{\nonumber\\}

\def\mm{\mathcal}

\def\pa{\partial}

\definecolor{db}{rgb}{0,0.08,0.45}
\definecolor{brick}{rgb}{0.6,0.1,0.3}

\definecolor{zz}{rgb}{1,0,0}
\definecolor{zz2}{rgb}{0.7,0.1,0.1}
\definecolor{yy}{rgb}{0.05,0.9,0.05}
\definecolor{ww}{rgb}{0.6,0.1,0.3}

\definecolor{rr}{cmyk}{0,0,0,1}
\definecolor{vv}{rgb}{0.5,0,0.5}
\definecolor{ss}{cmyk}{0,0,0,1}

\definecolor{brick}{rgb}{0.5,0,0.5}

\def\a{\alpha} \def\b{\beta} \def\g{\gamma} \def\d{\delta} \def\e{\epsilon}   \def\n{\nu} \def\m{\mu} \def\s{\sigma} \def\r{\rho}   \def\l{\lambda}  
\def\G{\Gamma}  \def\D{\Delta}     \def\S{\Sigma} \def\O{\Omega}

\title{On the Shape Dependence of Entanglement Entropy}

\author{Dean Carmi}

\affiliation{Raymond and Beverly Sackler Faculty of Exact Sciences School of Physics and Astronomy Tel-Aviv University, Ramat-Aviv 69978, Israel}

\emailAdd{deancarmi1@gmail.com}

\abstract{	We study the shape dependence of entanglement entropy (EE) by deforming symmetric entangling surfaces. We show that entangling surfaces with a rotational or translational symmetry extremize (locally) the EE with respect to shape deformations that break some of the symmetry (i.e. the 1st order correction vanishes). This result applies to EE and Renyi entropy for any QFT in any dimension. Using Solodukhin's formula in $4d$ and holography in any $d$, we calculate the 2nd order correction to the universal EE for CFTs and simple symmetric entangling surfaces. In all cases we find that the 2nd order correction is positive, and thus the corresponding symmetric entangling surface is a local minimum. Some of the results are extended to free massive fields and to 4d Renyi entropy.}

\begin{document}
	
	\maketitle

\section{Introduction}

Entanglement entropy is a measure of the quantum correlations of a system. It has a very wide range of applications from condensed matter physics to quantum field theory \cite{Casini:2009sr,Calabrese:2009qy,Calabrese:2004eu,Calabrese:2005zw,Solodukhin:2011gn,Casini:2011kv,Holzhey:1994we,Larsen:1995ax,Bombelli:1986rw,Callan:1994py,Srednicki:1993im,LevinWen,KitPres05,Casini:2012ei,Casini:2004bw,Hertzberg:2010uv,Lewkowycz:2012qr,Huerta:2011qi,Hung:2011ta,Fur13,Park:2015dia,Ben-Ami:2015zsa,Casini:2014yca,Liu:2012eea,Liu:2013una,Nozaki:2014uaa,Herzog:2014fra,Goykhman:2015sga,Casini:2015woa,Elvang:2015jpa,Donnelly:2015hxa,Zhou:2015cpa,Herzog:2015cxa} \nocite{Carmi:2011dt}
 and AdS/CFT \cite{Ryu:2006ef,Ryu:2006bv,Nishioka:2009un,Lewkowycz:2013nqa,Myers:2010tj,Myers:2010xs,Faulkner:2013yia,Hartman:2013mia,Schwimmer:2008yh,Faulkner:2013ica,Blanco:2013joa,Headrick:2007km,Kol:2014nqa,Agon:2015mja,Ghoroku:2015apa,Bhattacharya:2014vja,Nozaki:2013wia,Nozaki:2013vta,Georgiou:2015pia,He:2014lfa,Chakraborty:2014lfa,Faulkner:2014jva,Parnachev:2015nca,Czech:2015qta,Momeni:2015laa,Momeni:2015vka,Bueno:2015rda,Bueno:2015xda}. The EE in QFT is generally hard to calculate, and most computations have been done in simple setups, such as:  free fields, CFTs, and symmetric entangling surfaces (e.g. spheres and planes). Thus there is a need to obtain analytical results for interacting theories, non-CFTs, and less symmetrical entangling surfaces. This work aims to make a step in this direction by studying the shape dependence of EE. Previous works on the shape dependence of EE include \cite{Allais:2014ata,Mezei:2014zla,solo,Rosenhaus:2014woa,Rosenhaus:2014nha,Rosenhaus:2014ula,Rosenhaus:2014zza,Lewkowycz:2014jia,Klebanov:2012yf,Banerjee:2011mg,Myers:2013lva,Fonda:2014cca,Nakaguchi:2014pha,Ben-Ami:2014gsa,Huang:2015bna,Astaneh:2014uba,Fursaev:2012mp,Safdi:2012sn,Miao:2015iba}.

The divergent structure of entanglement entropy for a CFT in $d$-dimensions is:
\bea
\label{eq:div6kk} 
S = c_{d-2}\frac{R^{d-2}}{\d^{d-2}} + c_{d-4}\frac{R^{d-4}}{\d^{d-4}} + \ldots + 
\begin{Bmatrix}
	c_1\frac{R}{\d} +(-1)^{\frac{d-1}{2}}S^{(univ)}\ \ \ \ \ \ \ , \ \ \ \ d=odd \\ 	c_2 \frac{R^2}{\d^2} +(-1)^{\frac{d-2}{2}}S^{(univ)}\log(\frac{R}{\d})  \ \ \ , \ \ \ d=even
\end{Bmatrix}
\eea 
where $R$ is the scale of the entangling region, and $\d$ is the UV cutoff. The leading divergence is the area law, and all of the power law divergences are non-universal. We will be interested in the universal term $S^{(univ)}$, which in $d=even$ is the coefficient of the log divergence and in $d=odd$ it is the finite term. 

\begin{figure}
	\centering
	
	\begin{minipage}{0.48\textwidth}
		\centering
		
		\includegraphics[width=50mm]{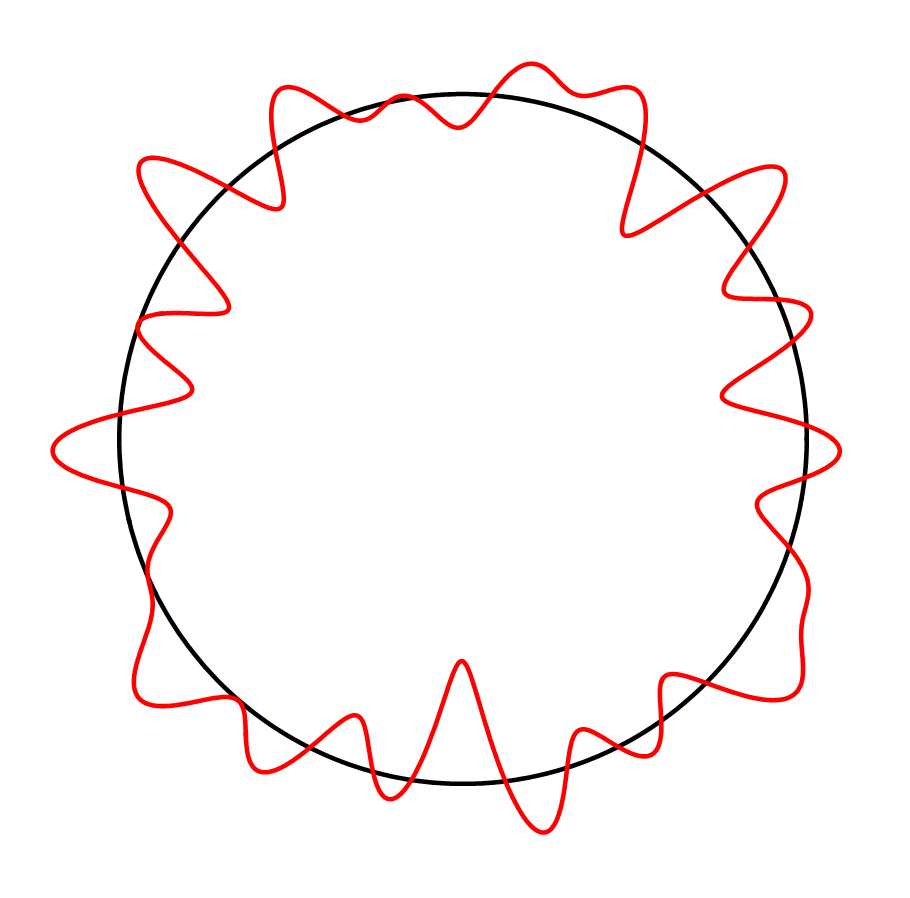}
	\end{minipage}
	\begin{minipage}{0.48\textwidth}
		\centering
		
		\includegraphics[width= 50mm]{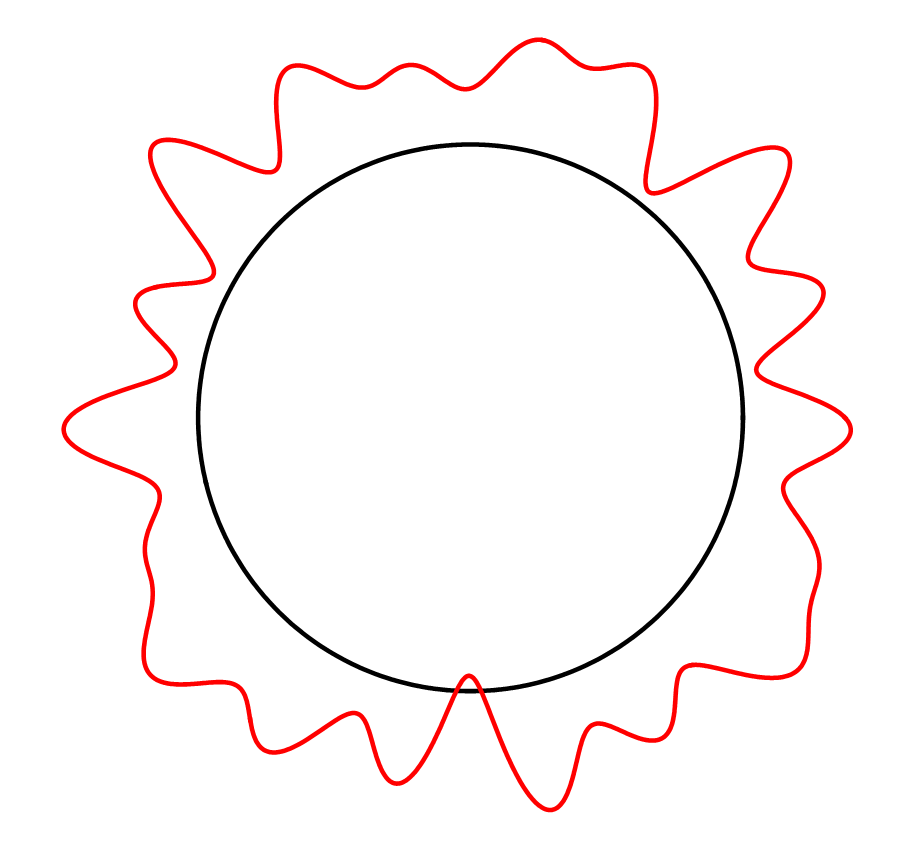}
	\end{minipage}
	\caption{Illustration of a perturbed circle $r(\phi)= 1 +\e \sum_n a_n \cos(n\phi)$.  \textbf{Left:} A Perturbation without a zero mode $a_0 =0$. \textbf{Right:} A Perturbation with a zero mode $a_0 =0.3$. \label{dd91}}
\end{figure}

Consider a QFT parametrized by coordinates ($t, y_i$,$r$), where $i=1\ldots, d-2$. For instance take $r$ to be the radial coordinate in spherical coordinates, and $y_i$ to be angles parameterizing the entangling surface. We will always work in a constant time slice $t=0$.
Consider a codimension-2 entangling surface defined by:
\bea
r(y_i)= r_0(y_i)
\eea 
where $r_0(y_i)$ is some given function of $y_i$. Thus we choose $r$ to be the dependent coordinate, and $y_i$ as the independent coordinates. We denote the entanglement entropy corresponding to the entangling surface $r_0(y_i)$ as $S_0$.
Now we slightly perturb the entangling surface:
\bea
\label{eq:polimnert6sgkl}
r(y_i)= r_0(y_i) +\e f(y_i)
\eea 
where $\e$ is a small parameter, and $f(y_i)$ is some arbitrary perturbation function\footnote{As an example, a perturbed circle in $d=3$ is shown in Fig.~\ref{dd91}. In this case, (\ref{eq:polimnert6sgkl}) is given by: $r(\phi)= R[1 +\e \sum_n a_n \cos(n\phi)]$, where we Fourier expanded the perturbation $f$. $R$ is the radius of the circle, $\phi$ is the angle in polar coordinates, and $a_n$ are the Fourier coefficients.}.


The entanglement entropy will change as a result of the perturbation of the entangling surface, and it can generally be written as an expansion in $\e$: 
\bea
\label{eq:polmg3}
S= S_0 + S_1 \e+S_2\e^2 + \ldots
\eea 

The above procedure was carried out in \cite{Allais:2014ata,Mezei:2014zla} for the case of a perturbed sphere for a CFT in $d$ dimensions. They start with a sphere entangling surface in a flat space-time background, and perturb the sphere as follows:
\bea
\label{eq:polimnescv4}
r(\O_{d-2})= R\Big[1 +\e  \sum_{l,m_1, \dots m_{d-3}} a_{l,m_1, \dots m_{d-3}} Y_{l,m_1, \dots m_{d-3}}(\O_{d-2})\Big]
\eea 
where $R$ is the sphere radius, the $a$'s are constants, and the $Y$'s are (real) hyper-spherical harmonics.
Then they calculate the resulting change in the universal term of the EE. They find that the the change in the universal EE vanishes at $1^{st}$ order, namely

\bea
S_1^{(univ)}=0
\eea 

Thus, for a CFT the sphere is a local\footnote{If the topology of the entangling surface is allowed to change, then the EE can become unbounded from below, as shown in 4d in \cite{Perlmutter:2015vma}.} extremum with respect to perturbations of the entangling surface. Additionally, for holographic CFTs \cite{Mezei:2014zla} uses the Ryu-Takayanagi formula \cite{Ryu:2006ef,Ryu:2006bv} (more specifically, its generalization to higher derivative gravity \cite{Dong:2013qoa,Camps:2013zua,Bhattacharyya:2013jma,Bhattacharyya:2013gra}) to calculate the $2^{nd}$ order correction $S_2^{(univ)}$:
\bea
\label{eq:kam}
S^{(univ)}_2= C_T \frac{\pi^{\frac{d+2}{2}}(d-1)}{2^{d-2}\G(d+2)\G(\frac{d}{2})} \sum_{l,m_1 \ldots m_{d-1}} a^2_{l,m_1 \ldots m_{d-1}}   \prod_{k=1}^d(l+k-2) \times
\begin{Bmatrix}
	\frac{\pi}{2} \ \ \ , \ \ \ d=odd \\ 1  \ \ \ , \ \ \ d=even
\end{Bmatrix} \nn
\eea 
where $C_T$ is the (positive) central charge appearing in the 2-point function: $\langle TT \rangle \sim \frac{C_T}{x^{2d}}$. A priori, $S^{(univ)}_2$ could have depended on the three parameters $C_t, t_2$, and $t_4$ of the 3-point function $\langle TTT \rangle$, but it turns out that it depends just on $C_T$.

$S^{(univ)}_2$ is clearly positive, implying that the sphere is a local minimum for holographic CFTs. (\ref{eq:kam}) was compared to $S^{(univ)}_2$ obtained from Solodukhin's formula in 4d CFTs, and precise agreement was found.

\begin{figure}
	\centering
	
	\begin{minipage}{0.48\textwidth}
		\centering
		
		\includegraphics[width=50mm]{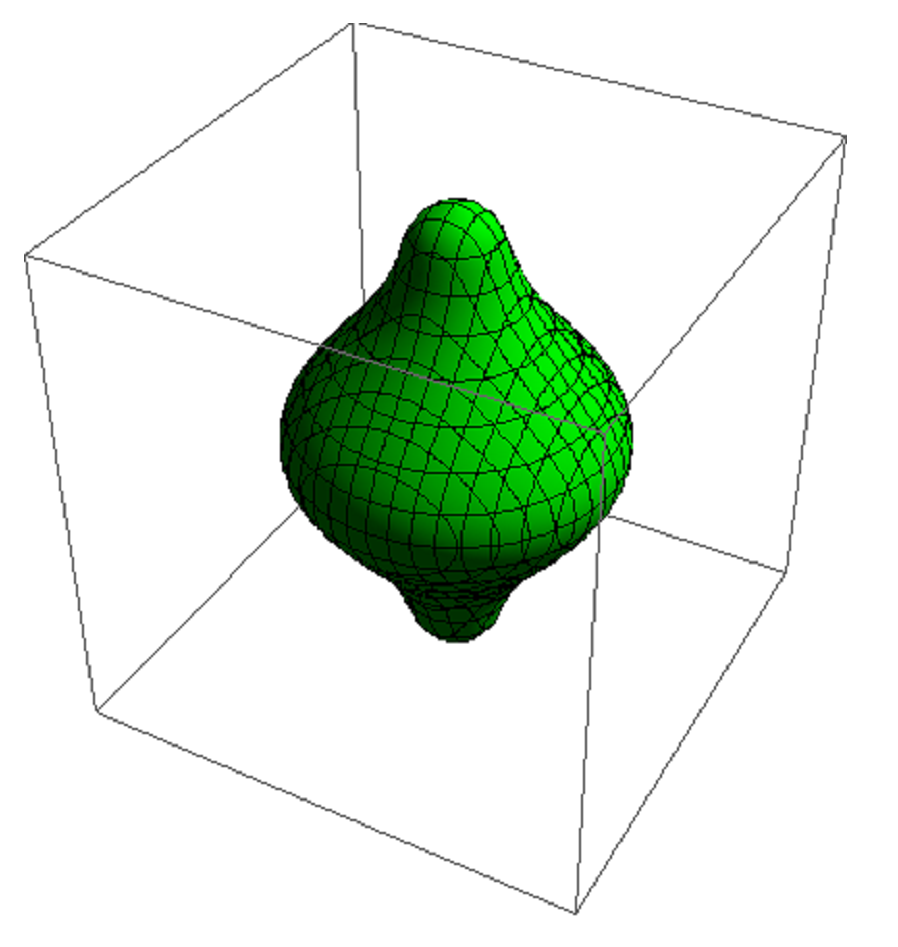}
	\end{minipage}
	\begin{minipage}{0.48\textwidth}
		\centering
		
		\includegraphics[width= 25mm]{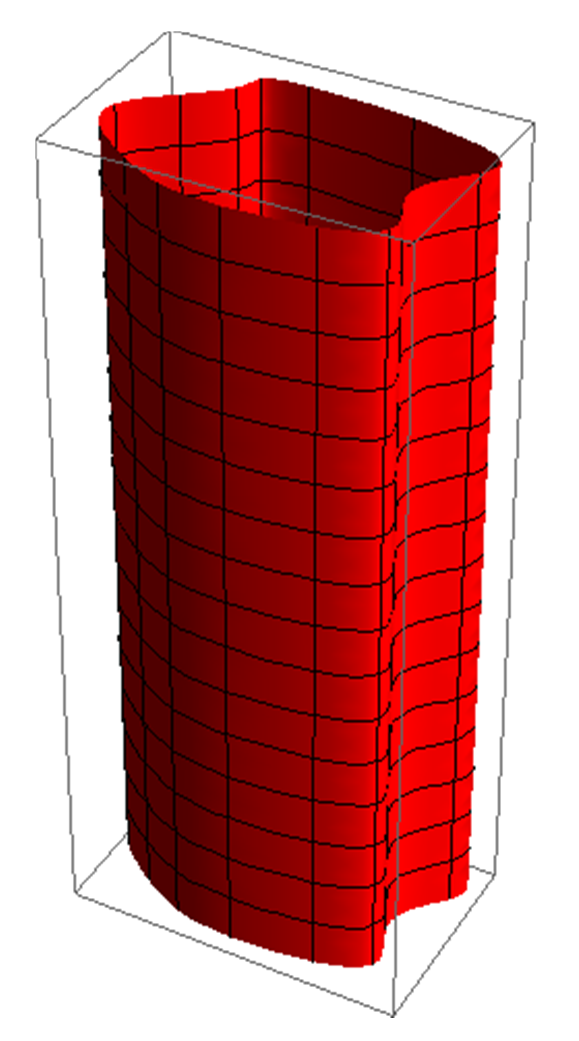}
	\end{minipage}
	\caption{\textbf{Left:} An example of a $d=4$ surface of revolution entangling surface with rotational symmetry. \textbf{Right:} A $d=4$ waveguide entangling surface with translational symmetry. Such symmetric entangling surfaces can obviously be generalized to higher dimensions.  \label{dd1}}
\end{figure}

In this work we will generalize the above results of  \cite{Mezei:2014zla} to less symmetric entangling surfaces, and (in some cases) to non-CFTs.
We will show that entangling surfaces with a rotational or translational symmetry in some direction (see Fig.~\ref{dd1}), extremize the universal EE with respect to shape deformations that break some of the symmetry\footnote{It should be emphasized that the family of shape perturbations that extremize the EE, are only those which break some of the symmetry of the original entangling surface. For example, for a 3d circle the perturbation in Fig.~\ref{dd91}-Left is allowed, whereas the perturbation in Fig.~\ref{dd91}-Right is not allowed because it has a component in the radial direction (a zero mode of the Fourier expansion).  Such perturbations cannot deform a sphere into a larger/smaller sphere. More generally, such perturbations cannot deform a surface of revolution into a different surface of revolution (containing the same number of symmetries).}, i.e:
\bea
\label{eq:yuter66}
S_1^{(univ)}=0
\eea 
The proof of this result will be purely geometrical, and hence will apply to any QFT (and also to Renyi entropy). A simple corollary is that for $d=even$, (\ref{eq:yuter66}) is true also for multiply connected entangling surfaces\footnote{For $d=even$ the universal term of the EE is a log divergence which is determined locally by the shape of the entangling surface. Thus for a multiply connected entangling surface the log term is a superposition of the contribution from each separate piece of the entangling surface.}.

Additionally, we will calculate the $2^{nd}$ order correction $S^{(univ)}_2$ (whose sign determines if the local extremum is a minimum or a maximum) for some simple entangling surfaces. We perform this calculation using holography (the Ryu-Takayanagi formula), and also by using Solodukhin's formula (for 4d CFTs). In all of the the examples of symmetric entangling surfaces that we have checked, the $2^{nd}$ order correction is positive, and this corresponds to a local minimum. We conjecture this to hold more generally to symmetric entangling surfaces. We also comment on results for free massive fields, and 4d Renyi entropy.


\section{The First Order Correction: Stationarity}

\label{sec:lkop8}


Consider a QFT parametrized by coordinates ($\phi$, $y_i$, $r$), where $i=1\ldots, d-3$. Assume $\phi$ to be a symmetry direction of the entangling surface.
Now perturb the entangling surface (see (\ref{eq:polimnert6sgkl})) with a single Fourier mode\footnote{A general perturbation (without a zero mode) can be written as:
	\bea
	r(\phi,y_i)=  r_0(y_i) +\e \sum_{n \neq 0} \Big ( A_n a_n(y_i) \cos (n\phi) +   B_n b_n(y_i) \sin (n\phi) \Big) 
	\eea 
	Then the EE can be expanded:
	\bea
	S= S_0 + \e S_1 +  \e^2 S_2 + \ldots
	\eea 
	
	At linear order in $\e$ the modes don't mix, and their contributions to $S_1$ add up linearly:
	\bea
	S_1 = \sum_{n \neq 0} \Big (A_n U_1(n) +  B_nU_2(n) \Big) 
	\eea 
	where  $U_1(n)$ and $U_2(n)$ are some functions of $n$. 
	Because the modes don't mix at linear order, we can compute the contribution of a single mode and then sum over all modes.
	
} :

\bea
\label{eq:sdc7}
r(\phi,y_i)=  r_0(y_i)  +\e  A_n a_n(y_i) \cos (n\phi)   \ \ \ \ , \ \ \ \  n\neq 0 
\eea 
where $r_0(y_i)$ doesn't depend on the symmetry direction $\phi$, the $a_n(y_i)$ are functions of $y_i$, and $A_n$ are constants.
The resulting EE can be expanded as:

\bea
\label{eq:poit4}
S= S_0 + \e S_1 +  \e^2 S_2 + \ldots
\eea 

Now lets consider the same perturbation but with negative sign, i.e. $\e \to -\e$:
\bea
\label{eq:sdc8}
\tilde{r}(\phi,y_i)=  r_0(y_i)  - \e   A_n a_n(y_i) \cos (n\phi)   \ \ \ \ , \ \ \ \  n\neq 0 
\eea

The resulting EE will be (just flipping the sign of $\e$ in (\ref{eq:poit4}):
\bea
\label{eq:rfvx5}
\tilde{S}= S_0 - \e S_1 + \e^2 S_2 + \ldots
\eea 

But the two perturbations (\ref{eq:sdc8}) and (\ref{eq:sdc7}) describe precisely the same entangling surface, only rotated. This can be seen by performing $\phi \to \phi + \frac{\pi}{n}$ on (\ref{eq:sdc8}), which gives (\ref{eq:sdc7}). Since the two entangling surfaces are the same, they have same EE and therefore from (\ref{eq:poit4}) and (\ref{eq:rfvx5}) we have:
\bea
\tilde{S} = S \ \ \ \ \ \ \ \ \ \longrightarrow \ \ \ \ \ \ \ \ \  S_1=0
\eea 

and we proved what we wanted.\\

An important point to make is that the proof above used only the rotation symmetry, and did not use any specific property of entanglement entropy or of the QFT. Therefore stationarity will hold for any quantity which is a function of the surface (e.g. Renyi entropy), and for any QFT.


\section{The Second Order Correction in Holography}

\label{sec:holog1}

Consider a boundary QFT parametrized by coordinates ($t, y_i, r$), $i=1\ldots, d-2$. $z$ is the holographic coordinate, and we set $t=0$.
The holographic EE according to the Ryu-Takayanagi formula is:
\bea
\label{eq:rtg6}
S= \int   d^{d-2}y_i dz \  \mm{L}(z,r,y_i)
\eea 
where $\mm{L}\equiv \frac{\sqrt{\det g}}{4G_N} $. The corresponding equation of motion for the bulk surface is:
\bea
\label{eq:rtg7}
\label{eq:eomlk5}
\frac{\pa \mm{L}}{\pa r} - \frac{d}{dz}\frac{\pa \mm{L}}{\pa (\pa_z r)} -  \frac{d}{dy_i}\frac{\pa \mm{L}}{\pa (\pa_{y_i} r)}  =0
\eea 
where there is a summation convention on $y_i$.

Consider an entangling surface defined by:
\bea
r(y_i)= r_0(y_i)
\eea 

Now perturb the entangling surface:
\bea
\label{eq:polimnert6}
r(y_i)= r_0(y_i) +\e f(y_i)
\eea 

where $\e$ is a small parameter.
Then the bulk surface will also get perturbed\footnote{We use the same letter $r$ for the entangling surface and the bulk surface. They can simply be distinguished by the fact that for the bulk surface there is a $z$ dependence.}:
\bea
\label{eq:4ndkdjj44}
r(z,y_i)= r_0(z,y_i) +\e r_1(z,y_i) + \e^2r_2(z,y_i) + \ldots
\eea


For the rest of this section we assume $d=even$ dimensions, and we will want to compute the universal log term.
We can then write the resulting perturbed $\mm{L}$ and $S$ as: 
\bea
S= S_0 + S_1 \e+S_2\e^2 + \ldots =  \int d^{d-2}y_i  dz \mm{L}= \int d^{d-2}y_i  dz \Big[ \mm{L}_0 + \mm{L}_1 \e+\mm{L}_2\e^2 + \ldots  \Big]
\nn
\eea 

We can derive a formula for the $2^{nd}$ order correction $S_2$. Assuming that the entangling surface has a symmetry in all directions (e.g. the cylinder surface in the next section), terms containing $r_2$ will fall and we get:

 \bea
\label{eq:seconddraw} 
S_2= \frac{1}{2} \Bigg\{  \int d^{d-2}y_i dz \bigg[  
r_1   \frac{d}{d\e} \frac{d}{dy_i}\frac{\pa \mm{L}}{\pa (\pa_{y_i} r)}    + \pa_{y_i} r_1  \frac{d}{d\e} \frac{\pa \mm{L}}{\pa (\pa_{y_i} r)}   \bigg]_{\e=0}   + 
\int_{\pa \mm{M}} d^{d-2}y_i    r_1 \frac{d}{d\e} \frac{\pa \mm{L}}{\pa (\pa_z r)} \bigg|_{\e=0}   \bigg|_{z=\d}^{z_{max}}  \Bigg\} 
\nn   
\eea 

\subsection{Cylinder Entangling Surface}

\label{sec:cyliner5g}

Now let us assume that the entangling surface is a cylinder $S^p \times \mathbb{R}^{d-2-p}$ in flat space-time. We denote $y_j$ as cartesian directions along $\mathbb{R}^{d-2-p}$ (where $-L\leq y_j \leq L$) and $\Omega_p$ as directions along the sphere $S^p$ (of radius $R$).
The holographic EE written in these cylindrical coordinates is (see also (\ref{eq:lkmf6})):
\bea
\label{eq:lk5mf6}
S= \int   dz d^{d-2-p}y_j d\Omega_p\  \mm{L}(z,r,y_j, \Omega_p)
\eea 

where:
\bea
\label{eq:lkmf6999}
\mm{L}(z,r,y_i, \Omega_p)= \frac{r^p F_1(z)}{z^{d-1}}\sqrt{1+ F_2(z)(\pa_z r)^2+ F_3(z)\Big[(\pa_{y_j} r)^2+\frac{1}{r^2}(\pa_{\Omega_p} r)^2\Big]}
\eea 
We are not committing yet to a particular metric in the bulk, therefore $F_1(z)$, $F_2(z)$, $F_3(z)$ are arbitrary functions of $z$. For an AdS metric we have: $F_2(z)=F_3(z)=1$, and $F_1=\frac{R^{d-1}_{AdS}}{4G_N}$.\\


We Fourier expand the perturbation $f$ in (\ref{eq:polimnert6}):
\bea
\label{eq:cyleom88912}
f(y_j,\Omega_p)=   \sum_{\{  n_j, l_p \}}   \Big[ a_{\{  n_j, l_p \}}  Y_{l_p}(\Omega_p)\prod_{i=1}^{d-2-p} \cos (n_i y_i) \Big]
\eea 
where $Y_{l_p}(\Omega_p)$ are (real) hyperspherical harmonics, the $n_j$  are integers, the $a_{\{  n_j, l_p \}}$ are the coefficients of the Fourier expansion, and for conciseness we defined $\{  n_j \} \equiv n_1,\ldots, n_{d-2-p}$  and $\{  l_p \} \equiv l,m_1 \ldots, m_{p-1}$. 

The bulk surface perturbation $r_1$ of (\ref{eq:4ndkdjj44}) can also be Fourier expanded (see (\ref{eq:cyleom8})):
\bea
\label{eq:cyleom889}
r_1(z,y_j,\Omega_p)=   \sum_{\{  n_j, l_p \}}   \Big[ a_{\{  n_j, l_p \}} r^{(1)}_{\{  n_j, l_p \}}(z) Y_{l_p}(\Omega_p)\prod^{d-2-p}_{i=1} \cos (n_i y_i) \Big]
\eea 
where $r^{(1)}_{\{  n_j, l_p \}}(z)$ are functions of $z$, which must obey the boundary condition: $r^{(1)}_{\{  n_j, l_p \}}(z=0)=1$.

Now we can derive an expression for the boundary term (the last term on the RHS of (\ref{eq:seconddraw})):
\bea
\label{eq:sn66sjk}
S_2^{{bound.}} =
\frac{L^{d-2-p}F_1(z) F_2(z)r_0^{p-1}}{2z^{d-1}[1+F_2(\pa_z r_0)^2]^{1/2}}   \sum_{\{  n_j, l_p \}}    a_{\{  n_j, l_p \}}^2 \bigg[\frac{r_0 r^{(1)}_{\{  n_j, l_p \}}\pa_zr^{(1)}_{\{  n_j, l_p \}}}{1+F_2(\pa_z r_0)^2 }  + p  \pa_z r_0 (r^{(1)}_{\{  n_j, l_p \}})^2 \bigg] \bigg|_{z=\d}
\nn
\eea
where $L$ is the length of the $\mathbb{R}^{d-2-p}$ directions, and we take the limit $L\to \infty$. This formula is a boundary term which is evaluated at the boundary $z=\d$. We will want to extract the universal log divergence from this formula in $d=even$ dimensions.

In principle, $r_0(z)$ and $r^{(1)}_{\{  n_j, l_p \}}(z)$ can be obtained by solving the EOM's for the bulk minimal surface (\ref{eq:sefjkddj}) and (\ref{eq:cylmmd7}). 
These solutions will generally have Fefferman-Graham type expansions of the form (\cite{Fefferman:2007rka,Graham:1999pm,Schwimmer:2008yh,Hung:2011ta}):
\bea
\label{eq:re7jfjj4}
r_0(z)= q_0 + q_2 z^2 + \ldots + q_{d}z^{d} + \tilde{q}_{d}z^{d} \log (z) + \ldots
\eea 

\bea
\label{eq:re7jfjj41st}
r^{(1)}_{\{  n_j, l_p \}}(z) = u_0 + u_2z^2 + \ldots + u_{d}z^{d} + \tilde{u}_{d}z^{d} \log (z) + \ldots
\eea

The terms in (\ref{eq:sn66sjk}) can then be written as:
\bea
\frac{r_0 r^{(1)}_{\{  n_j, l_p \}}\pa_zr^{(1)}_{\{  n_j, l_p \}}}{[1+F_2(\pa_z r_0)^2]z^{d-1}}\Big|_{\log} =d q_0 u_0 \tilde{u}_d \log z
\ \ \ \ \   ,  \ \ \ \ \ \ \  
\frac{p  \pa_z r_0 (r^{(1)}_{\{  n_j, l_p \}})^2}{z^{d-1}}\Big|_{\log} =dp u_0^2 \tilde{q}_d  \log z
\nn
\eea 




 
 
 where we extracted the log terms. Using (\ref{eq:sn66sjk}), and also the fact that the background is asymptotically AdS: $F_1(0)=\frac{R^{d-1}_{AdS}}{4G_N}=\frac{2 \pi^{\frac{d+2}{2}}(d-1)\G(\frac{d}{2})}{\G(d+2)}C_T$ and $F_2(0)=1$, we get\footnote{Note that for a CFT, dimensional analysis (and (\ref{eq:sefjkddj}), (\ref{eq:cylmmd7})) dictate the functional dependence of  $\tilde{q}_d$ and $\tilde{u}_d$ such that: $\tilde{q}_d = f_1(d)R^{-d+1}$ and $\tilde{u}_d = f_2(l,\tilde{n}, R,d)$, where $f_1$, $f_2$ are some functions and $\tilde{n}^2 \equiv \sum_i n_i^2$. Note that there is no dependence on $m_1, \ldots , m_{p-1}$, since the bulk EOM's (\ref{eq:sefjkddj}), (\ref{eq:cylmmd7}) do not depend on them. Likewise it can be seen that the sphere result (\ref{eq:kam}) doesn't depend on the $m$'s.}
 \bea
 \label{eq:lkcylind5}
 S^{bound.}_2\big|_{\log} =  \frac{(-1)^{\frac{d-2}{2}} \pi^{\frac{d+2}{2}}\G(\frac{d}{2})}{(d+1)\G(d-1)}C_T  L^{d-2-p} R^{p-1}  \sum_{\{  n_j, l_p \}}  a_{\{  n_j, l_p \}}^2  \Big(R \tilde{u}_d  +p  \tilde{q}_d\Big)  \log \big(\frac{R}{\d}\big)
 \nn
 \eea 
where we used $u_0=1$ and $q_0=R$, and multiplied by $(-1)^{\frac{d-2}{2}}$ , see (\ref{eq:div6kk}).

It can be shown that for the two special cases $p=0,1$, the boundary term above is the \underline{only} contribution to $S^{(univ)}_2$. Thus the $p=1$ case gives: 
\bea
\label{eq:lkcylind55}
S^{(univ)}_2 =  \frac{(-1)^{\frac{d-2}{2}} \pi^{\frac{d+2}{2}}\G(\frac{d}{2})}{(d+1)\G(d-1)}C_T  L^{d-3}   \sum_{\{  n_j, l \}}  a_{\{  n_j, l \}}^2  \Big(R \tilde{u}_d  +  \tilde{q}_d\Big) 
\nn
\eea 
We see from (\ref{eq:lkcylind55}) that the sign of $S^{(univ)}_2$ depends solely on the coefficients $\tilde{u}_{d}$ and $\tilde{q}_{d}$ of the log term in the FG expansions. These coefficients can be obtained by solving the bulk EOMs  (\ref{eq:sefjkddj}) and (\ref{eq:cylmmd7}). It might be interesting to understand if generally these coefficients are constrained to have a definite sign.
The $p=0$ case (the plane) is examined in the following subsection.

\subsection{Plane Entangling Surface}

The plane entangling surface $\mathbb{R}^{d-2}$ is a special case of the cylinder with $p=0$. Therefore (\ref{eq:lkcylind5}) becomes\footnote{Note that (\ref{eq:lkcylind5}), (\ref{eq:lkcylind55}), and (\ref{eq:lkcylind599}) apply to any asymptotically AdS background.}:

\bea
\label{eq:lkcylind599}
	S^{(univ)}_2 =  \frac{(-1)^{\frac{d-2}{2}} \pi^{\frac{d+2}{2}}\G(\frac{d}{2})}{(d+1)\G(d-1)}C_T  L^{d-2}   \sum_{\{  n_j\}}  a_{\{  n_j \}}^2   \tilde{u}_d  
\nn
\eea 

So the sign of $S^{(univ)}_2$ depends solely on the coefficient $\tilde{u}_{d}$, which we shall now determine.  
For a plane entangling surface and Einstein gravity in the bulk, we will find an explicit solution\footnote{This result was also derived in \cite{Nozaki:2013vta} in the context of "entanglement density".} (\ref{eq:9jkgh3}):
\bea
\label{eq:sumifff}
x^{(1)}_{\{  n_i\}}(z)=  \frac{1}{\mm{N}}z^{\frac{d}{2}}K_{\frac{d}{2}}(\tilde{n} z)
\eea 
where $\mm{N}$ is a normalization constant, and $\tilde{n}^2 \equiv \sum_i n_i^2$.
The small $z$ expansion of this function gives (see (\ref{eq:bessl99})) $\tilde{u}_d = \frac{(-1)^{\frac{d-2}{2}}\tilde{n}^{d}}{2^{d-2}d(\frac{d}{2}-1)!^2}$.   Therefore we have:

\bea
\label{eq:planefinla}
	S^{(univ)}_2=  
	 \frac{\pi^{\frac{d-2}{2}}(d-1)}{2^{d-2}\G(d+2)(\frac{d}{2}-1)!} C_T L^{d-2}   \sum_{\{  n_j\}}^{\infty}   \tilde{n}^d a_{\{  n_j\}}^2    
\nn
\eea

Since the above expression is positive, we have proved that a plane is a local minimum for Einstein gravity in the bulk. It is now natural to conjecture that (\ref{eq:planefinla}) holds for any CFT. (\ref{eq:planefinla}) was derived for $d=even$, but it would be simple task to obtain the analogous $d=odd$ result. The $d=odd$ result will differ from (\ref{eq:planefinla}) only by its $d$ dependence, and for $d=3$ it was obtained in \cite{Huang:2015bna}.

For the case $d=4$, (\ref{eq:planefinla}) gives:
\bea
\label{eq:hologhj88}
	S_2^{(univ)}=  C_T\frac{\pi^3L^2}{160}    \sum_{n_2=1}^{\infty}\sum_{n_3=1}^{\infty}   (n_2^2+n_3^2)^2 a_{n_2,n_3}^2   
\nn
\eea

This precisely matches (\ref{eq:vfvfvf43432}) which we will obtain in the next section for 4d CFTs via Solodukhin's formula.


\section{The Second Order Correction in Field theory}

\subsection{4d CFT: Solodukhin's Formula}

\label{sec:4dsolop}

For a $d=even$ CFT there is a universal log term (\ref{eq:div6kk}):
\bea
S\big|_{\log} = (-1)^{\frac{d-2}{2}}S^{(univ)}\log \big(\frac{R}{\d}\big)
\eea 
and as in (\ref{eq:polmg3}), we can expand in small $\e$:
\bea
\label{eq:polmg3h}
S^{(univ)} = \sum_{k=0}^{\infty}  S^{(univ)}_k \e^k
\eea 

Solodukhin's formula \cite{solo} (which applies to a CFT in 4d) is\footnote{$a_4$ and $c_4$ are the a and c anomalies in 4d. Both are normalized such that for a real scalar field their value is 1.  We will sometimes use $C_T=\frac{c_4}{3\pi^4}$ instead of $c_4$.}:
\bea
\label{eq:soldfgh}
S^{(univ)}= \frac{a_4}{180}\int_{\S}d^2 \s \sqrt{\g}E_2 + \frac{c_4}{240 \pi}\int_{\S}d^2 \s \sqrt{\g}I_2
\eea 
where $E_2$ is the Euler density, and the integrals are over the entangling surface $\S$, and
\bea
\label{eq:soldfghy6}
I_2\equiv Tr ( k^2) - \frac{1}{2}k^ak^a = (k^a_{\m \n}-\frac{1}{2} \g_{\m \n}k^a)^2 \geq 0
\eea 
 
The second fundamental form and extrinsic curvature are defined as:
\bea
k_{\m \n}^a = \g^\a_\m \g^\b_\n \nabla_\a \hat{n}^a_\b    \ \ \ \ \ , \ \ \ \ \ \   k^a= Tr (k^a_{\m \n}) = \g^{\m \n}k^a_{\m \n}   \ \ \ \ \ , \ \ \ \ \ \  Tr ( k^2) = \g^{\m \n} \g^{\r \s}k^a_{\n \r} k^a_{\s \m} 
\nn
\eea 

where $\g_{\m \n}$ is the induced metric on the entangling surface.

We consider shape perturbations that leave the topology of the entangling surface fixed, therefore the change in the Euler term above is zero, and $S_2^{(univ)}$ of (\ref{eq:polmg3h}) will be given by the integral of $I_2$:
\bea
\label{eq:soldfgh3j}
S^{(univ)}_2 = \frac{c_4}{240 \pi}\int_{\S}d^2 \s \sqrt{\g}I_2 \Big|_{\e^2}
\eea 

In (\ref{eq:soldfghy6}) $I_2$ is always positive \cite{Astaneh:2014uba,Perlmutter:2015vma}  and is zero for the sphere. Therefore the sphere locally minimizes $I_2$ and $S^{(univ)}$. This argument also works for flat entangling surfaces (plane, strips) since these have $I_2=0$. 

In the following, we calculate $S^{(univ)}_2$ for several examples. See also \cite{Allais:2014ata,Mezei:2014zla}.\\

$\bullet$\ \  \textbf{Example 1: Plane Entangling Surface}\\

The metric in cartesian coordinates is:
\bea
ds^2 = dx^2 + dy_2^2 + dy_3^2
\eea

Consider a plane at $x=0$, where $y_2$, $y_3$ are coordinates along the surface. Now perturb it as follows:
\bea
x= 0 +\e  f(y_2, y_3)
\eea

The vector normal to the surface is:
\bea
\hat{n}_\m = \frac{1}{\sqrt{1+\e^2 (f^2_{y_2}+f^2_{y_3})} } (1\ , -\e f_{y_2} , -\e f_{y_3} ) \ \ \ \ \ , \ \ \ \ \ \ \text{where}\ \ \ \ \ \ \ \ \ f_{y_i} \equiv \pa_{y_i} f
\eea

The second fundamental form at order $O(\e^2)$ is:
\begin{eqnarray}
k_{\m \n}^a=
\begin{bmatrix}
0 & -\e^2(f_{y_3}f_{y_2 y_3}+ f_{y_2}f_{y_2 y_2}) & -\e^2(f_{y_3}f_{y_3 y_3}+ f_{y_2}f_{y_2 y_3})\\
-\e^2(f_{y_3}f_{y_2 y_3}+ f_{y_2}f_{y_2 y_2})& -\e  f_{y_2 y_2} & -\e  f_{y_2 y_3}\\
-\e^2(f_{y_3}f_{y_3 y_3}+ f_{y_2}f_{y_2 y_3})&  -\e  f_{y_2 y_3} &  -\e  f_{y_3 y_3} \\
\end{bmatrix}
\nn
\end{eqnarray}

$k_{\m \n}^a$ starts at order $\e$ since the unperturbed plane is flat. Plugging this in (\ref{eq:soldfghy6}) gives:
\bea
\label{eq:finaly7}
\sqrt{\g}I_2= I_2+ O(\e^3)= \frac{\e^2}{2} \Big[(f_{y_2 y_2}-f_{y_3 y_3})^2 +4f^2_{y_2y_3}  \Big] (f^2_{y_2}+f^3_{y_3}) + O(\e^3)
\eea 
Note that $I_2$ starts at order $\e^2$, therefore the order $\e$ correction vanishes even before integration over the surface.  
Now Fourier expand the perturbation: 

\bea 
\label{eq:finaly74f}
f(y_2, y_3)=  \sum_{n_2,n_3=1}^{\infty} a_{n_2 n_3}  \cos (n_2 y_2)  \cos (n_3 y_3)
\eea
and plug (\ref{eq:finaly7}), (\ref{eq:finaly74f}) in  (\ref{eq:soldfgh3j}):
\bea 
\label{eq:vfvfvf43432}
	S^{(univ)}_2\ =   C_T \frac{\pi^3L^2}{160} \sum_{n_2=1}^{\infty}\sum_{n_3=1}^{\infty} (n_2^2+n_3^2)^2 a_{n_2n_3}^2     
\eea	
where we used $c_4=3\pi^4C_T$, and the integral $\int_{-L}^{L}dy \cos^2 (ny)=L$, where the width of the plane is very large $L\to \infty$.
So we got a positive result, and therefore the universal term of a plane in 4d is a local minimum. We see that (\ref{eq:vfvfvf43432}) precisely matches the result (\ref{eq:hologhj88}) obtained in holography.

We can compute higher orders of $\e$ in (\ref{eq:polmg3h}), and we note that for the plane all odd terms vanish: $S^{(univ)}_{2k+1}=0$. At $4^{th}$ order we get:
\bea 
S^{(univ)}_4 =  - C_T\frac{\pi^3L^2}{11520} \sum_{n_2=1}^{\infty}\sum_{n_3=1}^{\infty} a_{n_2 n_3}^4 \Big(15n_2^6+ 15n_3^6 +13n_2^4n_3^2+13n_3^4n_2^2 \Big) 
\eea
\\


$\bullet$\ \  \textbf{Example 2: Sphere Entangling Surface}\\

This case was calculated in \cite{Allais:2014ata,Mezei:2014zla}, and we write their result.

The perturbed sphere entangling surface is:
\bea
r(\theta,\phi)= R[1 + \e f(\theta,\phi)]  = R\Big[1 + \e \sum_{l,m} a_{lm}Y_{lm}(\theta, \phi)\Big]  
\eea 
where $R$ is the radius of the sphere.
Plugging in (\ref{eq:soldfgh3j}) gives:
\bea
S^{(univ)}_2 =  C_T \frac{\pi^3}{160} \sum_{l,m}a_{lm}^2  (l-1)l(l+1)(l+2)
\eea 

This matches the holographic result (\ref{eq:kam}) \cite{Mezei:2014zla}.\\


$\bullet$\ \   \textbf{Example 3: Cylinder Entangling Surface}\\

The metric in cylindrical coordinates is:
\bea
ds^2 = dr^2 + dy^2 + r^2  d\phi^2
\eea

Consider a perturbed cylinder entangling surface with radius $R$:
\bea
r(y, \phi)= R +\e f(y, \phi)
\eea

We get:
\bea
\sqrt{\g}I_2 \big|_{\e^2}=  
\frac{1}{4R^3}[ 2f^2 + 3f_\phi^2  + 2f_{\phi \phi}^2 +8f f_{\phi \phi} -R^2 f_y^2 +8R^2 f^2_{y\phi} -4 R^2 f_{yy}f_{\phi \phi} +2R^4 f_{yy}^2   ]\e^2
\nn
\eea 

Plugging $f(y,\phi) =  \sum_{n,m}a_{n,m} \cos (m \phi) \cos (n y)$,
\bea
\label{eq:ybkkgp76}
S_2^{(univ)}=  \frac{c_4}{240 \pi}\int_{\S}dy d\phi \sqrt{\g}I_2 \Big|_{\e^2} =
\nn
C_T \frac{\pi^4L}{320R^3} \sum_{n,m}a_{nm}^2 \Big[ 2 -5m^2  +2m^4  + (nR)^2(4m^2-1) +2 (nR)^4  \Big]
\eea 
Thus $S_2^{(univ)}\geq 0$ for $m\geq 2$ and for all $n$, and the cylinder is a local minimum.\\


$\bullet$\ \  \textbf{Example 4: 4d Surface of Revolution}\\

Consider a surface of revolution entangling surface (e.g. Fig.~\ref{dd1}-Left) given by:
\bea
r(\theta,\phi)= r_0(\theta)
\eea 
where we use spherical coordinates $(r,\theta, \phi)$.
This surface has rotational symmetry in the $\phi$ direction. Now perturb the surface as follows:
\bea
\label{eq:medna1}
r(\theta,\phi)= r_0(\theta)[1+\e f_2(\theta)\cdot f_3(\phi)]
\eea 

with $f_3(\phi) = \sum_{m\neq 0} a_m \cos (m\phi)$. (\ref{eq:soldfgh}) and (\ref{eq:soldfghy6}) give at most 4 derivatives of $\phi$, thus the result can be written as a polynomial in $m$ (after integrating over $\phi$):
\bea
\label{eq:finish2}
S^{(univ)}_2 =  \frac{c_4}{240 \pi}\int_{\S}d\theta d \phi \sqrt{\g}I_2  \Big|_{\e^2} = 
\nn
\frac{\pi^3 C_T}{80}\int d\theta \sum_{m\neq 0} a_m^2\Big[G_1(r_0, f_2, \theta)m^4 +  G_2(r_0, f_2, \theta)m^2 +  G_3(r_0, f_2, \theta) \Big]
\eea 
where the $G_i(r_0, f_2,\theta)$ are some functions of $r_0,f_2$ and their derivatives. Explicit calculation gives the $m^4$ coefficient:

\bea
G_1(r_0, f_2, \theta) =
\frac{ \pi f_2^2\Big(4+4\csc^2(\theta)r_0^2r_0'^8 + 16\csc^2(\theta)r_0^4r_0'^6 + 24\csc^2(\theta)r_0^6r_0'^4 + 16r_0^8r_0'^2 \Big)}{8r_0^3\sin \theta(r_0^2+r_0'^2)^{\frac{9}{2}}}
\nn
\eea 
Since $r_0\geq 0 $, we see that $G_1(r_0, f_2, \theta) \geq 0$. Therefore for perturbations with large enough $m$ (short wave-length perturbations) (\ref{eq:finish2}) is positive, and thus all 4d surfaces of revolution are local minima. 
We are not able to show that the functions $G_2(r_0, f_2, \theta)$, $G_3(r_0, f_2, \theta)$ are positive (though the final integrated result $S^{(univ)}_2$ may still turn out to be positive).\\


$\bullet$\ \  \textbf{Example 5: 4d Waveguide Surface}\\

Consider a general waveguide entangling surface (e.g. Fig.~\ref{dd1}-Right) given by:
\bea
r(y, \phi)= r_0(\phi)
\eea 
where we use cylindrical coordinates $(r, y, \phi)$.
This surface has translational symmetry in the $y$ direction. Now we perturb the surface as follows:
\bea
\label{eq:medna17}
r= r_0(\phi)[1+\e f_2(\phi) f_3(y)]
\eea 

with $f_3(y) = \sum_{m\neq 0}  a_m \cos (my)$. The result can be written as a polynomial in $m$ (after integrating over $y$):
\bea
\label{eq:finish1}
S^{(univ)}_2=  \frac{c_4}{240 \pi}\int_{\S}d\phi d y \sqrt{\g}I_2  \Big|_{\e^2} = 
\nn
\frac{\pi^3 C_T}{80}\int d\phi \sum_{m\neq 0} a_m^2\Big[G_1(r_0, f_2)m^4 +  G_2(r_0, f_2)m^2 +  G_3(r_0, f_2) \Big]
\eea 
where the $G_i(r_0, f_2)$ are some functions of $r_0,f_2$ and their derivatives. 
Explicit calculation gives the $m^4$ coefficient:

\bea
G_1(r_0, f_2) =
\frac{ \pi f_2^2\Big(2r_0^{10}+6r_0^8r_0'^2 +6r_0^6r_0'^4+2r_0^4r_0'^6 \Big)}{4(r_0^2+r_0'^2)^{\frac{7}{2}}} \geq 0
\nn
\eea 
Therefore for perturbations with large enough $m$ (short wave-length perturbations) (\ref{eq:finish1}) is positive, and thus all 4d waveguide surfaces are local minima. 
We are not able to show that the functions  $G_2(r_0, f_2)$, $G_3(r_0, f_2)$ are positive (though the final integrated result $S^{(univ)}_2$ may still turn out to be positive).

\subsection{Renyi Entropy}

There is a  generalization of (\ref{eq:soldfgh}) to Renyi entropy in a 4d a CFT \cite{Fursaev:2012mp},\cite{Lewkowycz:2014jia}:
\bea
\label{eq:renyi887}
\mm{S}_q^{(univ)}= Q_1(q)\frac{a_4}{180}\int_{\S}d^2 \s \sqrt{\g}E_2 + Q_2(q)\frac{c_4}{240 \pi}\int_{\S}d^2 \s \sqrt{\g}I_2 
\nn
\eea
where $\mm{S}_q^{(univ)}$ is the universal log term for the $q$-th Renyi entropy. $Q_{1,2}(q)$ are functions of $q$ such that $Q_1(1)=Q_2(1)=1$, so it matches (\ref{eq:soldfgh}) in the EE limit $q\to 1$. Up to these functions, (\ref{eq:renyi887}) and (\ref{eq:soldfgh}) are the same, hence the results in the previous sections can be used. In particular, if $Q_2(q)$ is positive then for any entangling surface the Renyi entropy will have the same sign as the EE. Note that $Q_2(q)$ is positive for free fields, and there is strong evidence that it is positive for holographic CFTs \cite{Lewkowycz:2014jia,Lee:2014xwa,Perlmutter:2015vma}. If this turns out to be correct, then the Renyi entropy will be a local minimum whenever the EE is.


\subsection{Free Massive Fields}

Certain universal EE terms for free massive field theories have been found \cite{Hertzberg:2010uv,Lewkowycz:2012qr,Huerta:2011qi,Hung:2011ta}. The so called "universal area law" for free scalars or fermions with mass $m$ has the following form:  
\bea
\label{eq:sign66}
S=
\begin{Bmatrix}
	(-1)^\# \gamma_d A_{\S} m^{d-2}  \ \ \ \ \ \ , \ \ \  \ \ \ d=odd \\ 		(-1)^\# \tilde{\gamma}_d A_{\S} m^{d-2} \log (m\d)  \ \ \ , \ \ \ d=even
\end{Bmatrix}
\eea 
where $\gamma_d$ is a positive constant that depends only on the dimension $d$. All of the shape dependence is contained in $A_{\S}$, the area of the entangling surface.
Table~\ref{tbl:ledkk} lists the sign factors $(-1)^\#$ in (\ref{eq:sign66}) for a Dirac fermion, a conformally coupled scalar, and a minimally coupled scalar.

\begin{table}[h]
	\centering
	\begin{tabular}{|c|c|c|c|}
		\hline
		& Dirac Fermion   & Conformal scalar & Minimal scalar\\ 
		\hline
		$d=odd$		& $(-1)^{\frac{d-1}{2}}$ &  $(-1)^{\frac{d+1}{2}}$ & $(-1)^{\frac{d-1}{2}}$  \\  
		\hline
		$d=even$		&$(-1)^{\frac{d}{2}}$ & $(-1)^{\frac{d-2}{2}}$  & $(-1)^{\frac{d}{2}}$ \\  
		\hline
	\end{tabular}
	\caption{The sign factor in the "universal area law" (\ref{eq:sign66}). \label{tbl:ledkk}} 
\end{table}

As an example, let us now compute the "universal area law" term for a deformed sphere entangling surface:
\bea
\label{eq:dolmit67}
r(\Omega_{d-2})=  R\Big[1 + \e \sum_{\{lm\}}    a_{\{lm\}}  Y_{\{lm\}}(\Omega_{d-2}) \Big]
\eea 

The area of the deformed sphere is:
\bea
A_\S = \int d\O_{d-2} r^{d-2} =\frac{(d-1)\pi^{\frac{d-1}{2}}}{\G(\frac{d+1}{2})}R^{d-2} + \e^2 \frac{(d-2)(d-3)}{2} R^{d-2}\sum_{\{lm\}}   a_{\{lm\}}^2 + O(\e^3)
\nn
\eea 
where we plugged (\ref{eq:dolmit67}) and performed the integrals.

The first term on the RHS is the area of the undeformed sphere, the $O(\e)$ correction vanishes as expected, and the $O(\e^2)$ correction is positive. Therefore the $O(\e^2)$ correction in $S$ has the same sign as the zeroth order, which can be read from Table~\ref{tbl:ledkk}. 
Generalizing to non-spheres, it is easy to see that the area of a perturbed surface of revolution is larger than that of the unperturbed surface of revolution. It would be interesting to perform a similar analysis to higher curvature terms for free massive fields (i.e to curvature corrections to the "universal area law"). 

A similar analysis applies to "universal area law" terms in interacting theories \cite{Hung:2011ta,Rosenhaus:2014zza,Park:2015dia}, with a shape dependence that comes only from the area $A_\S$. The EE for a CFT perturbed by a relevent operator of dimension $\D=\frac{d+2}{2}$, contains the following term:
\bea
S=  N\l^2 \frac{d-2}{4(d-1)}\frac{\pi^{\frac{d+2}{2}}}{\G(\frac{d+2}{2})}A_{\S} \log \big(\frac{R}{\d}\big) 
\eea 
where $\l$ is the coupling constant. Such log terms occur both in odd and even dimensions.


\section{Discussion}

In this work we studied the shape dependence of entanglement entropy by deforming symmetric entangling surfaces. We showed that entangling surfaces with a rotational or translational symmetry locally extremize the EE with respect to shape deformations that break some of the symmetry. This result applies to EE and Renyi entropy for any QFT in  any dimension. Using Solodukhin's formula and holography, we calculated the 2nd order correction to the EE for CFTs and simple symmetric entangling surfaces. In all cases we found that the 2nd order correction is positive, and thus the corresponding symmetric entangling surface is a local minimum. Perhaps this result holds more generally for any symmetric entangling surface.

Let us mention some possible future directions.

\begin{itemize}
\item The calculation in section \ref{sec:holog1} considered only Einstein gravity in the bulk, and it would be interesting to consider also higher derivative gravity. For spheres, \cite{Mezei:2014zla} found that $S_2^{(univ)}$ depends just on $C_T$ and not on $t_2$ or $t_4$ (these are the three parameters in the 3-point function of stress tensors). It would be nice to check if this continues to hold for other entangling surfaces.

\item Computing the FG coefficients $\tilde{u}_d$ and $\tilde{q}_d$ in (\ref{eq:lkcylind5}) by solving the bulk EOMs for the cylinder (\ref{eq:sefjkddj}) and (\ref{eq:cylmmd7}). Maybe it is also possible to show, using a more general principle, that $\tilde{q}_d$ and $\tilde{u}_d$ in (\ref{eq:lkcylind55}) must have a definite sign. It would also be interesting to consider (in holography) more general entangling surfaces with a symmetry.

\item The work of \cite{Rosenhaus:2014zza} attempted to compute $S_2^{univ}$ for a plane entangling surface in a $d=4$ CFT using the perturbative formalism of EE \cite{Rosenhaus:2014woa}. They were not able to obtain the $I_2$ term of Solodukhin's formula (\ref{eq:soldfgh}), and the current situation is somewhat puzzling. Our result (\ref{eq:planefinla}) as well as (\ref{eq:kam}) (obtained in \cite{Mezei:2014zla}) might help in resolving this puzzle.

\item It would be interesting to repeat the analysis of section~\ref{sec:4dsolop} for 6d CFTs using the results of \cite{Safdi:2012sn,Miao:2015iba}.

\item Other possible extensions are to compute higher orders $S_j^{(univ)}$ for $j>2$, and also to perform computations in a curved space-time background.
\end{itemize}

\newpage 

\textbf{Acknowledgments}

I thank Carlos Hoyos, Eric Perlmutter, Misha Smolkin, and especially Omer Ben-Ami for helpful discussions. I am very grateful to Zohar Komargodski, Mark Mezei, and Shimon Yankielowicz for valuable discussions and collaboration during some stages of this work. Our work is partially supported by the Israel Science Foundation (grant 1989/14), the US-Israel bi-national fund (BSF) grant 2012383 and the German Israel bi-national fund GIF grant number I-244-303.7-2013.


\begin{appendix}

\section{Equations of Motion for the Minimal Bulk Surface}
	
In this section, we obtain EOMs for the bulk minimal surfaces of cylinder, strip, and plane. We solve the $1^{st}$ EOM for the plane, and obtain $\tilde{u}_d$ in (\ref{eq:lkcylind599}). 

\subsection{Strip Entangling Surface}

Consider a strip entangling surface and a bulk metric in Poincare coordinates:
\bea
ds^2= \frac{R^2_{AdS}}{z^2}\Big[\frac{1}{\b(z)}dz^2-dt^2 +dx^2 + dy_i^2\Big]
\eea
Where $\b(z)$ is some function of the holographic coordinate $z$, which for AdS: $\b(z)=1$. $x$ is the direction perpendicular to the strip entangling surface, $y_i$ are directions along the strip entangling surface, and $i=1, \dots , d-2$.
The holographic EE is (see (\ref{eq:rtg6}), (\ref{eq:rtg7})):
\bea 
S= \frac{R^{d-1}_{AdS}}{4G_N}\int d^{d-2} y_i  \int_{\d}^{z_{max}} dz   \frac{1}{z^{d-1}} \sqrt{\frac{1}{\b(z)}\Big[1+(\pa_{y_i} x)^2\Big] +(\pa_z x)^2}
\eea
	
In the ansatz of (\ref{eq:lkmf6999}) this corresponds to $p=0$, $F_1=\frac{R^{d-1}_{AdS}}{4G_N\sqrt{\b(z)}}$, $F_2=\b(z)$, and $F_3=1$. The corresponding EOM is:
\bea
\label{eq:stripool}
\frac{d}{dz}\Bigg(\frac{1}{z^{d-1}}\frac{\pa_z x}{ \sqrt{\frac{1}{\b(z)}\Big[1+(\pa_{y_i} x)^2\Big] +(\pa_z x)^2}}\Bigg) +
\frac{1}{z^{d-1}} \frac{d}{d{y_i}}\Bigg(\frac{\pa_{y_i} x}{ \sqrt{\frac{1}{\b(z)}\Big[1+(\pa_{y_i} x)^2\Big] +(\pa_z x)^2}}\Bigg)  =0
\nn
\eea 
	
The bulk surface can be expanded in $\e$:
\bea
\label{eq:stripoo}
x(z,y_i)= x_0(z) + \e x_1(z,y_i) + \e^2 x_2(z,y_i) + \ldots
\eea  
where because of the translational symmetry of the strip, $x_0(z)$ does not depend on $y_i$.
The EOM (\ref{eq:stripool}) at $0^{th}$ order in $\e$ is:
\bea
\frac{d}{dz} \bigg(  \frac{\pa_z x_0}{z^{d-1}\sqrt{\frac{1}{\b(z)}+(\pa_z x_0)^2}} \bigg)=0 
\eea 
	
The solution to this is:
\bea
\label{eq:zero9} 
(\pa_z x_0)^2 = \frac{z^{2d-2}}{\b(z)(z_{max}^{2d-2}-z^{2d-2})} 
\eea  
where $z_{max}$ is the turning point of the bulk surface.
The EOM at $1^{st}$ order in $\e$ is:
\bea
\label{eq:zero9h} 
\frac{d}{dz}\Bigg(\frac{1}{z^{d-1}}\frac{\frac{1}{\b(z)}\pa_z x_1}{ \big(\frac{1}{\b(z)}+(\pa_z x_0)^2\big)^{3/2} }\Bigg) +
\frac{1}{z^{d-1}} \frac{\pa^2_{y_i} x_1}{ \sqrt{\frac{1}{\b(z)}+(\pa_z x_0)^2 }}  =0
\eea 
	
Plugging (\ref{eq:zero9}) in (\ref{eq:zero9h}):
\bea
\frac{d}{dz}\Bigg(\frac{\b^{\frac{1}{2}}}{z^{d-1}}\frac{(z_{max}^{2d-2}-z^{2d-2})^{3/2}\pa_z x_1}{ z_{max}^{2d-2} }\Bigg) +
\frac{\b^{\frac{1}{2}}}{z^{d-1}}  (z_{max}^{2d-2}-z^{2d-2})^{1/2} \pa^2_{y_i} x_1  =0
\eea 
	
Simplifying, we get:
\bea
\label{eq:stripbbb}
\pa_z^2 x_1 + \Bigg[  \frac{\b'(z)}{2\b(z)}- \frac{(d-1)}{z}\cdot \frac{ 1+ 2\Big(\frac{z}{z_{max}}\Big)^{2d-2} }{ 1- \Big(\frac{z}{z_{max}}\Big)^{2d-2} }\Bigg]\pa_z x_1  + \frac{\pa_{y_i}^2x_1}{ 1- \Big(\frac{z}{z_{max}}\Big)^{2d-2}  }=0
\eea 
	
This equation seems hard to solve analytically. In the next section we consider the simpler case of a plane entangling surface.

\subsection{Plane Entangling Surface}

For a plane entangling surface situated at $x=0$, the bulk minimal surface goes straight down in the bulk: $x_0(z)=0$. We have (see (\ref{eq:stripoo})): 
\bea
x(z,y_i)= \e x_1(z,y_i) + \e^2 x_2(z,y_i) + \ldots
\eea

The turning point of the bulk surface is at $z_{max} \to \infty$, therefore (\ref{eq:stripbbb}) becomes:
\bea
\label{eq:nb6h3n}
\pa_z^2 x_1 +\bigg[\frac{\b'(z)}{2\b(z)}- \frac{(d-1)}{z}\bigg]\pa_z x_1 + \pa_{y_i}^2 x_1=0
\eea 
	
We Fourier expand $x_1$:
\bea
x_1(z,y)=  \sum_{\{  n_i \}}   a_{\{  n_i \}} x^{(1)}_{\{  n_i \}}(z) \prod^{d-2}_{i=1} \cos (n_i y_i) 
\eea 
where the $n_i$ are integers, and we defined the shorthand notation: $\{  n_i \} \equiv n_1,\ldots, n_{d-2}$.
Plugging this in (\ref{eq:nb6h3n}) gives:
\bea
\label{eq:xcjgg} 
\pa_z^2 x^{(1)}_{\{  n_i \}} + \bigg[\frac{\b'(z)}{2\b(z)}- \frac{(d-1)}{z}\bigg]\pa_z x^{(1)}_{\{  n_i \}} - \tilde{n}^2 x^{(1)}_{\{  n_i \}} = 0
\eea 
where we defined $\tilde{n}^2 = \sum_{i} n_i^2$. 
Now we consider the ansatz $\b(z)= 1+ \a z^k$ for the bulk metric, then:
\bea
\pa_z^2 x^{(1)}_{\{  n_i \}} + \Big[\frac{\a kz^{k-1}}{2(1+ \a z^k)} - \frac{d-1}{z}\Big]\pa_z x^{(1)}_{\{  n_i \}} -  \tilde{n}^2 x^{(1)}_{\{  n_i \}} = 0
\eea 
	
The CFT case $\b(z)=1$ can be recovered by plugging $\a=0$:
\bea
\label{eq:plnf56}
\pa_z^2 x^{(1)}_{\{  n_i \}} -  \frac{d-1}{z} \pa_z x^{(1)}_{\{  n_i \}} - \tilde{n}^2 x^{(1)}_{\{  n_i \}} =0
\eea

This equation is solved by modified Bessel functions\footnote{ The modified Bessel function:
\bea
\label{eq:bessl99}
K_\n (z)= \frac{1}{2}\Big(\frac{z}{2}\Big)^{-\n}\sum_{j=0}^{\nu-1} (-1)^j\frac{(\nu -j-1)!}{4^{j} j!} z^{2j}\  +\ (-1)^{\nu+1}\log(z/2)I_\nu (z)
\nn
+\ (-1)^\n \frac{1}{2}\Big(\frac{z}{2}\Big)^{\n} \sum_{j=0}^\infty \frac{\psi (j+1) + \psi(n+j+1)}{ 4^j j!(\n+j)!} z^{2j}
\eea 	
	
where $\psi$ is the digamma function, and
\bea
\label{eq:bessl9934}
I_\n (z)= \Big(\frac{z}{2}\Big)^{\n} \sum_{j=0}^{\infty} \frac{1}{4^{j} j! \G(\n+j+1)} z^{2j} 
\eea 
}:
 
\bea
x^{(1)}_{\{  n_i \}}(z) =  C_1 z^{\frac{d}{2}}K_{\frac{d}{2}}(\tilde{n}z) + C_2 z^{\frac{d}{2}}I_{\frac{d}{2}}(\tilde{n}z)
\eea 
	
Choosing a boundary condition such that $x^{(1)}_{\{  n_i \}}(z)$ does not explode at $z \to \infty$, leaves only $K_{\frac{d}{2}}$. Therefore the final solution is:
\bea
\label{eq:9jkgh3}
x^{(1)}_{\{  n_i \}}(z) =  \frac{1}{\mm{N}} z^{\frac{d}{2}}K_{\frac{d}{2}}(\tilde{n} z)
\eea 
where the normalization constant is $\mm{N}= 2^{\frac{d}{2}-1}(\frac{d}{2}-1)! \tilde{n}^{-\frac{d}{2}}$. We will use the above solution in (\ref{eq:sumifff}). This result was also derived in \cite{Nozaki:2013vta} in the context of "entanglement density".

\subsection{Cylinder Entangling Surface}

Consider a cylinder entangling surface $S^p \times \mathbb{R}^{d-2-p}$ in flat space-time $\mathbb{R}^{d}$.
The holographic EE in cylindrical coordinates is (see (\ref{eq:lk5mf6}),(\ref{eq:lkmf6999})):
\bea
\label{eq:lkmf6}
S= \frac{R^{d-1}_{AdS}}{4G_N}\int dz d^{d-2-p}y d\Omega_p \frac{r^p}{z^{d-1}}\sqrt{1+(\pa_z r)^2+(\pa_{y_j} r)^2+\frac{1}{r^2}(\pa_{\Omega_p} r)^2}
\eea 

where we assumed the $AdS_{d+1}$ metric.
The corresponding EOM is:
\bea
\frac{d}{dz}\Bigg(\frac{r^p}{z^{d-1}}\frac{\pa_z r}{ \sqrt{1+(\pa_z r)^2+(\pa_{y_j} r)^2+\frac{1}{r^2}(\pa_{\Omega_p} r)^2}}\Bigg) + \frac{1}{z^{d-1}}\frac{d}{dy_i}\Bigg( \frac{r^p \pa_{y_i} r}{ \sqrt{1+(\pa_z r)^2+(\pa_{y_j} r)^2+\frac{1}{r^2}(\pa_{\Omega_p} r)^2}}\Bigg) 
\nn
\frac{1}{z^{d-1}} \frac{d}{d{\Omega_p}}\Bigg(\frac{r^{p-2} \pa_{\Omega_p} r}{ \sqrt{1+(\pa_z r)^2+(\pa_{y_j} r)^2+\frac{1}{r^2}(\pa_{\Omega_p} r)^2}}\Bigg) - \frac{1}{z^{d-1}} \frac{d}{dr}\Bigg(r^p \sqrt{1+(\pa_z r)^2+(\pa_{y_j} r)^2+\frac{1}{r^2}(\pa_{\Omega_p} r)^2}\Bigg) =0
\nn
\eea

The bulk surface expanded around the $0^{th}$ order cylinder is:
\bea
\label{eq:yy6dl4f}
r= r_0(z)+ \e r_1(z,y_j,\Omega_p)+\e^2 r_2(z,y_j,\Omega_p)+ \ldots
\eea

The $0^{th}$ order EOM is:
\bea
\label{eq:sefjkddj}
\frac{d}{dz}\Bigg(\frac{r_0^p}{z^{d-1}}\frac{\pa_z r_0}{ \sqrt{1+(\pa_z r_0)^2}}\Bigg) - \frac{1}{z^{d-1}} pr_0^{p-1} \sqrt{1+(\pa_z r_0)^2} =0
\eea

The $1^{st}$ order EOM is:
\bea
\label{eq:sefjkddj1st}
\pa^2_z r_1 +
\Bigg[  \frac{\frac{d}{dz} \Big( \frac{r_0^p}{z^{d-1}}\frac{1}{(1+(\pa_z r_0)^2)^{3/2}} \Big)}{\Big( \frac{r_0^p}{z^{d-1}}\frac{1}{(1+(\pa_z r_0)^2)^{3/2}} \Big) } \Bigg] \pa_z r_1 
+ \big[1+(\pa_z r_0)^2\big]\Big( \frac{p}{r_0^2}r_1 + \frac{1}{r_0^2}\pa_{\Omega_p}^2 r_1 + \pa_{y_j}^2 r_1     \Big) =0 
\nn
\eea

We Fourier expand $r_1$ (see (\ref{eq:cyleom889})):
\bea
\label{eq:cyleom8}
r_1(z,y_j,\Omega_p)=   \sum_{\{  n_j, l_p \}}   \Big[ a_{\{  n_j, l_p \}} r^{(1)}_{\{  n_j , l_p\}}(z) Y_{l_p}(\Omega_p)\prod_{i=1}^{d-2-p} \cos (n_i y_i) \Big]
\eea 

where the $n_j$  are integers, and we defined the shorthand notation: $\{  n_j \} \equiv n_1,\ldots, n_{d-2-p}$  and $\{  l_p \} \equiv l,m_1 \ldots, m_{p-1}$.
We plug (\ref{eq:cyleom8}) in (\ref{eq:sefjkddj1st}):
\bea
\label{eq:cylmmd7}
\pa^2_z r^{(1)}_{\{  n_j , l_p\}} +
\Bigg[  \frac{\frac{d}{dz} \Big( \frac{r_0^p}{z^{d-1}}\frac{1}{(1+(\pa_z r_0)^2)^{3/2}} \Big)}{\Big( \frac{r_0^p}{z^{d-1}}\frac{1}{(1+(\pa_z r_0)^2)^{3/2}} \Big) } \Bigg] \pa_z r^{(1)}_{\{  n_j , l_p\}} 
+ \Big[1+(\pa_z r_0)^2\Big]\Big( \frac{p-l(l+p-1)}{r_0^2} -\tilde{n}^2   \Big)r^{(1)}_{\{  n_j , l_p\}} =0 
\nn
\eea
where we defined $\tilde{n}^2\equiv \sum_j n_j^2$, and used $\pa^2_{\Omega_p}Y_{l_p} = -l(l+p-1) Y_{l_p}$ \cite{Mezei:2014zla}. Since the parameters $m_1, \ldots , m_{p-1}$ do not appear in (\ref{eq:cylmmd7}), the solution $r^{(1)}_{\{  n_j , l_p\}}$ will not depend on them (and neither will $S^{bound.}_2\big|_{\log}$ in (\ref{eq:lkcylind5})).

For a plane entangling surface we have $p=0$ and $r\equiv x$, and $\pa_z x_0 =0$, and in this case (\ref{eq:cylmmd7}) reduces to (\ref{eq:plnf56}).
On the other hand, for a sphere entangling surface we have $p=d-2$ and $r_0^2= R^2-z^2$. Thus for a sphere the EOM is:
\bea
\pa^2_z r^{(1)}_l 
- \frac{1}{z}\frac{(d-1)R^2+2z^2}{R^2-z^2}  \pa_z r^{(1)}_l
+ \frac{R^2[d-2-l(l+d-3)]}{(R^2-z^2)^2} r^{(1)}_l  =0 
\nn
\eea

We find a solution to this equation in terms of hypergeometric functions. For $d=3$ the solution is:
\bea
  r^{(1)}_l (z) = \frac{1}{\mm{N}}\Big(\frac{z-R}{z+R}\Big)^{\frac{l}{2}}\Big(\frac{R+lz}{\sqrt{z^2-R^2}}\Big)
\eea 
which agrees with Eq.~43 of \cite{Allais:2014ata}.




\end{appendix}

\bibliographystyle{utphys}

\bibliography{lib}

\end{document}